\RequirePackage{fix-cm}
\documentclass[smallextended]{svjour3}       % onecolumn (second format)
\smartqed  % flush right qed marks, e.g. at end of proof
\usepackage{graphicx}
\journalname{Hyperfine Interact (2012)}
\begin{document}

\title{CPT-symmetry studies with antihydrogen
}


\author{Ralf Lehnert}


\institute{Ralf Lehnert \at
              Indiana University Center for Spacetime Symmetries, 
Bloomington, IN 47405, USA\\
              Tel.: +1-812-855-6190\\
              Fax: +1-812-855-5533 \\
              \email{ralehner@indiana.edu} 
          }

\date{Published online: 24 January 2012}

\maketitle

\begin{abstract}
Various approaches to physics beyond the Standard Model can lead to small violations of CPT invariance. Since CPT symmetry can be measured with ultrahigh precision, CPT tests offer an interesting phenomenological avenue to search for underlying physics. We discuss this reasoning in more detail, comment on the connection between CPT and Lorentz invariance, and review how CPT breaking would affect the (anti)hydrogen spectrum.
\end{abstract}

\section{Introduction}
\label{intro}

Present-day established physics is well described by
two different, mutually incompatible frameworks: 
quantum theory, 
which governs nature at the microscopic level,
and classical general relativity, 
which dominates physical phenomena at large distance scales.
Although the description of the physical world by these two theories 
is tremendously successful phenomenologically,
the apparent necessity for two {\it distinct} frameworks 
is somewhat unsatisfactory from a theoretical perspective. 
For example, 
certain situations,
such as the description of the birth of our universe, 
are characterized by conditions that
call for the simultaneous application of both quantum theory and general relativity.

For this reason, 
significant efforts are presently directed towards 
a more fundamental description of nature that
contains both quantum theory and classical general relativity 
as limiting cases. 
While there are a number of theoretical approaches in this context,
phenomenological progress is hampered, 
primarily by the expected Planck suppression 
of quantum-gravity effects. 
A common avenue to tackle this issue is 
to look for model predictions that 
are {\em not} Planck suppressed. 
Examples are large extra dimensions 
or novel particles.

Another path towards quantum-gravity phenomenology 
is a bottom-up approach. 
It involves the identification of physical relations or physics principles
testable with present-day or near-future technology
at sensitivity levels that 
can be interpreted as having Planck reach. 
A further condition on such a relation or principle is that
it should hold {\em exactly} in established physics, 
so that even the smallest observed deviations 
would definitely imply new physics. 
The third desirable feature is that 
at least some theoretical approaches to quantum gravity
should allow for departures from the established physics in questions,
so as to provide motivation for research efforts along these lines.

Spacetime symmetries satisfy the above three requirements: 
they can be studied experimentally with ultra-high precision,
they hold exactly in known physics,
and various theoretical ideas in the field of quantum gravity 
can accommodate their breakdown~\cite{lotsoftheory}.
Spacetime symmetries therefore 
represent an excellent area for research 
in this context.

Spacetime symmetries fall into two classes:
continuous and discrete symmetries.
The ten continuous ones are 
four translations,
three rotations, 
and three Lorentz boosts.
The discrete spacetime symmetry
is CPT invariance.
Note that these symmetries are closely intertwined.
For example, 
in the context of axiomatic field theory 
with conventional quantum mechanics 
and local interactions,
CPT violation goes hand in hand with Lorentz breaking~\cite{AntiCPT}.

In what follows,
we will review the field-theory description 
of CPT violation at low energies
and discuss its predictions 
for (anti)hydrogen spectroscopy 
as well as antihydrogen free-fall measurements.

\section{Field theory for CPT violation}
\label{sme}

To test CPT invariance, 
it is desirable to employ a framework that
allows for departures from this symmetry: 
such a framework places the identification 
and analysis of suitable CPT tests 
on a more solid footing. 
Various levels of CPT test models 
ranging from {\em ad hoc} assumptions 
of mass differences between particle and antiparticle 
to effective field theory (EFT) can in principle be considered.
We proceed here at the level of EFT 
for the following reasons.
EFT is a tremendously successful tool in various areas of physics 
including elementary-particle,
nuclear, 
and condensed-matter physics. 
Note in particular that 
the latter area involves discrete backgrounds 
and non-relativistic dynamical aspects, 
features that
may have to be modeled also in quantum-gravity phenomenology.
Moreover, 
EFT can implement CPT violation at the most fundamental level 
in established physics,
which can guarantee internal theoretical consistency 
at all levels of presently known physics.
In particular, 
an EFT model can include the usual Standard Model and general relativity, 
so that essentially all presently feasible CPT tests can be studied.

Such an EFT framework for CPT violation---known as  
the Standard-Model Extension (SME)---is 
indeed available~\cite{SME}. 
As per the aforementioned ``Anti-CPT Theorem"~\cite{AntiCPT}, 
CPT- and Lorentz-breaking are closely linked in the SME. 
Note, 
however, 
that even though all  operators for departures from CPT invariance 
also violate Lorentz symmetry, 
the converse is not true.

The basic idea behind the inclusion of CPT and Lorentz breaking into the SME 
involves fixed, non-dynamical, tensorial structures that 
select preferred directions. 
More specifically, 
the SME effective Lagrangian ${\cal L}_{\rm SME}$ is
\begin{equation}
\label{smelagr}
{\cal L}_{\rm SME}={\cal L}_{\rm SM}+{\cal L}_{\rm EH}+\delta{\cal L}_{\rm SME}\,,
\end{equation}
where ${\cal L}_{\rm SM}$ and ${\cal L}_{\rm EH}$ 
denote the usual Standard-Model and Einstein--Hilbert terms, 
and $\delta{\cal L}_{\rm SME}$ contains small Lorentz- and CPT-violating contributions:
\begin{equation}
\label{LVterms}
\delta{\cal L}_{\rm SME}=
i\overline{\psi}c^{\mu\nu}\gamma_\mu\partial_\nu\psi-
\overline{\psi}a^{\mu}\gamma_\mu\psi+\ldots\,.
\end{equation}
Here, 
$c^{\mu\nu}$ and $a^{\mu}$ are SME coefficients,
where $a^{\mu}$ controls certain types of both Lorentz and CPT breakdown, 
while the Lorentz-violating $c^{\mu\nu}$ is CPT symmetric. 
Experimental studies seek to measure or constrain 
these (and other) tensorial coefficients.

The SME has been the framework for 
numerous experimental and phenomenological investigations of 
CPT and Lorentz invariance~\cite{ExpRev}, 
including ones involving cosmic radiation~\cite{UHECR},
particle colliders~\cite{collider},
resonance cavities~\cite{cavities},
neutrinos~\cite{neutrinos}, 
and precision spectroscopy~\cite{spectroscopy}.
A number of theoretical analyses have studied various aspects 
of the internal consistency of the SME~\cite{theory},
but no problematic issues have been found thus far.

\section{Tests with exotic atoms}
\label{EXA}

Exotic atoms provide an excellent test ground
for CPT and Lorentz symmetry for two reasons: 
the possibility for high-precision spectroscopy 
and the experimental access to types of matter 
difficult to study otherwise.
The prospect for comparing matter to antimatter, 
and the associated potential for ultra-sensitive CPT tests, 
is of particular interest in this context.
In what follows, 
we will focus on two specific classes of tests,
namely (anti)hydrogen spectroscopy 
and the interaction of antihydrogen with the gravitational field.

The effects of CPT and Lorentz violation 
on the spectrum of (anti)hydrogen 
within the minimal SME have been determined 
about a decade ago~\cite{AntiH} 
and have previously been discussed at EXA~\cite{EXAR}.
It is therefore sufficient 
to give just a brief description of these results 
in the present work.

A natural candidate transition for CPT tests in (anti)hydrogen
is the unmixed 1S--2S transition: 
the projected sensitivity 
is at the $10^{-18}$ level,
which is promising in the context of Planck-scale tests.
Unfortunately,
the highly symmetric composition of these states 
eliminates first-order effects in the minimal SME. 
From a theoretical perspective, 
this transition is therefore less useful for CPT tests.
Another option is a measurement of the spin-mixed 1S--2S transition.
A study within the minimal SME does show that
this transition would exhibit unsuppressed CPT violation.
But there may
be a practical disadvantage: 
the inhomogeneities in the trapping $B$ field  
would lead to a corresponding position dependence 
of the transition frequency, 
which represents a drawback for precision studies.
A third idea would be the experimental determination 
of the hyperfine Zeeman splitting within the 1S state itself. 
An analysis within the minimal SME indeed establishes
unsuppressed CPT-breaking effects, 
and in principle, 
a magnetic-field independent transition point 
can be selected.
Note in particular that 
similar transitions of this type, 
such as the usual Hydrogen-maser line,
can be resolved with ultra-high precision.

The EFT description of the interplay between gravity and CPT and Lorentz violation 
is more involved than
in the flat-spacetime limit.
For example, 
the Bianchi identities become nontrivial on a curved manifold 
and impose conditions on the CPT- and Lorentz-violating background.
One important theoretical result in this context is that
an explicitly symmetry-breaking background is usually inconsistent~\cite{SME}. 
A spontaneous violation of CPT and Lorentz invariance 
avoids such issues 
and is assumed in what follows.

For simplicity, 
we consider a sample scenario 
in which there is the only a single fermion of mass $m$, 
and the only non-zero SME coefficients are $a^\mu$ and $c^{\mu\nu}$. 
The corresponding terms in the Lagrangian are shown in Eq.~(\ref{LVterms}).
Our goal is now to extract the effective interaction of two point-like fermions 
in the non-relativistic limit for weak gravitational fields. 
This procedure requires various steps~\cite{IPM}. 
For instance, 
the aforementioned consistency condition of spontaneous symmetry breakdown
requires the initial introduction of a further dynamical field. 
The dynamics of these additional degrees of freedom 
must then be interpreted correctly 
(i.e., they must be integrated out). 

After various other considerations, 
and a further simplification to the iso\-trop\-ic case, 
only the two SME coefficients $a_T$ and $c_{TT}$ survive. 
The resulting two-parameter toy model is called the isotropic parachute model (IPM),
and its classical non-relativistic point-particle Lagrangian reads~\cite{IPM}
\begin{equation}
\label{ModNew}
L_{\rm IPM}=\frac{1}{2}
m_{\rm inertial
}
v^2+
\frac{
m_{\rm grav}M_{\rm grav}
}{r}\,,
\end{equation}
where the SME coefficients are contained in the effective inertial and gravitational masses: 
\begin{equation}
m_{\rm inertial}
=m(1+\frac{5}{3}
\overline{c}_{TT}
)\,,\qquad
m_{\rm grav}=m+
m
\overline{c}_{TT}
\pm
2
\alpha (\overline{a}_{\rm eff})_T\,.
\end{equation}
Here, 
$\overline{c}_{TT}$ and $(\overline{a}_{\rm eff})_T$ 
are effective coefficients for CPT and Lorentz breaking 
closely related to the original $a_T$ and $c_{TT}$. 
The $\pm$ sign distinguishes between fermion and antifermion.
The constant $\alpha$ 
depends on the details of the symmetry-violating field that
needed to be introduced for consistency. 

Lagrangian~(\ref{ModNew}) is reminiscent of 
the conventional Lagrangian for a classical non-relativistic point particle 
in the gravitational field of a second point particle.
All effects of CPT and Lorentz violation 
are absorbed into $m_{\rm inertial}$ and $m_{\rm grav}$. 
One can now consider various special regions 
in our toy model's two-dimensional 
$(m\overline{c}_{TT}, \alpha(\overline{a}_{\rm eff})_T)$ parameter space. 
For example, 
the special choice $\frac{1}{3}m\overline{c}_{TT}=\pm\alpha(\overline{a}_{\rm eff})_T$
means the following.
For the $+$ sign, 
the fermion behaves conventionally with $m_{\rm inertial}=m_{\rm grav}$, 
and free-fall modifications exist only for antifermions.
For the other sign choice, 
the situation would be vice versa: 
novel effects are confined to matter,
and antimatter behaves conventionally.
The IPM toy model therefore illustrates that
certain regions of the SME's parameter space 
are best tested with antimatter.

\begin{acknowledgements}
The author wishes to thank the organizers 
for arranging this stimulating meeting 
and for the invitation to participate.
This work is supported by the Indiana University Center for Spacetime Symmetries 
and the Portuguese Funda\c c\~ao para a Ci\^encia e a Tecnologia 
under Grant No.~CERN/FP/116373/2010.
\end{acknowledgements}


\end{document}